\documentclass[twocolumn]{aa} % for a paper on 1 column
\usepackage{graphicx}
\usepackage[varg]{txfonts}
\usepackage[percent]{overpic}
\usepackage{color}
\usepackage{multirow}
\usepackage{natbib,twoopt} % use author/date bibliographic citations
\usepackage[colorlinks=true,urlcolor=blue,linkcolor=red]{hyperref} %% to avoid \citeads line fills
\bibpunct{(}{)}{;}{a}{}{,} % to follow the A&A style
\begin{document}

   % To be published in A&A Section 13: Astronomical instrumentation
   \title{LARS -- An Absolute Reference Spectrograph for solar observations}
   \subtitle{Upgrade from a prototype to a turn-key system}

   \author{J. L\"ohner-B\"ottcher\inst{\ref{inst_kis}} \and W. Schmidt\inst{\ref{inst_kis}} \and H.-P. Doerr\inst{\ref{inst_mps},\ref{inst_kis}} \and T. Kentischer\inst{\ref{inst_kis}} \and T. Steinmetz\inst{\ref{inst_mpq},\ref{inst_menlo}} \and R. A. Probst\inst{\ref{inst_mpq},\ref{inst_menlo}} \and R. Holzwarth\inst{\ref{inst_mpq},\ref{inst_menlo}}}
    \institute{
    	Kiepenheuer-Institut f\"ur Sonnenphysik, Sch\"oneckstr. 6, 79104 Freiburg, Germany\\ \email{jlb@leibniz-kis.de}\label{inst_kis} \and
	Max-Planck-Institut f\"ur Sonnensystemforschung, Justus-von-Liebig-Weg 3, 37077 G\"ottingen, Germany\label{inst_mps} \and
	Max-Planck-Institut f\"ur Quantenoptik, Hans-Kopfermann-Strasse 1, 85748 Garching, Germany\label{inst_mpq} \and
	Menlo Systems GmbH, Am Klopferspitz 19, 82152 Martinsried, Germany\label{inst_menlo}}
   \date{Received 12 May 2017 / Accepted 04 July 2017}

  \abstract %max 300 words, self-consistent, no abbreviations
%Context
  {LARS is an Absolute Reference Spectrograph designed for ultra-precise solar observations. The high-resolution echelle spectrograph of the Vacuum Tower Telescope is supported by a state-of-the-art laser frequency comb to calibrate the solar spectrum on an absolute wavelength scale. In this article, we describe the scientific instrument and focus on the upgrades in the last two years to turn the prototype into a turn-key system.}
%Aims 
{The pursued goal was to improve the short-term and long-term stability of the systems, and enable a user-friendly and more versatile operation of the instrument.}
%Methods
  {The first upgrade involved the modernization of the frequency comb. The laser system generating the comb spectrum was renewed. The Fabry-P\'erot cavities were adjusted to filter to a repetition frequency of 8\,GHz. A technologically matured photonic crystal fiber was implemented for spectral broadening which simplified and stabilized the setup. The new control software facilitates an automated operation of the frequency comb. The second, quite recent upgrade was performed on the optics feeding the sunlight into a single-mode fiber connected to the spectrograph. A motorized translation stage was deployed to allow the automated selection of three different fields-of-view with diameters of 1\arcsec, 3\arcsec, and 10\arcsec\ for the analysis of the solar spectrum.}
%Results
  {The successful upgrades allow for long-term observations of up to several hours per day with a stable spectral accuracy of ${\rm 1\,m\,s^{-1}}$ limited by the spectrograph. The instrument covers a wavelength range between 480\,nm and 700\,nm in the visible. Stable, user-friendly operation of the instrument is supported. The selection of the pre-aligned fiber to change the field of view can now be done within seconds.}
%Conclusion
  {LARS offers the possibility to observe absolute wavelength positions of spectral lines and Doppler velocities in the solar atmosphere. First results demonstrate the capabilities of the instrument for solar science. The accurate measurement of the solar convection, p-modes, and atmospheric waves will enhance our knowledge of the solar atmosphere and its physical conditions to improve current atmospheric models.}
 % max 6 keywords
  \keywords{Instrumentation: spectrographs -- Instrumentation: miscellaneous -- Sun: photosphere -- Line: profiles}

  \maketitle
  \titlerunning{LARS -- An Absolute Reference Spectrograph for solar observations} 
  \authorrunning{L\"ohner-B\"ottcher et al.}

%##############################################################################################
%##############################################################################################
\section{Introduction}\label{sec_intro}
Observational solar physics relies strongly on precise and accurate spectroscopy. The optical solar spectrum with its thousands of spectral lines provides a wealth of information about material motions in the light-emitting layers of the Sun, the presence of magnetic fields, but also gas temperature, ionization state, etc. The Sun allows measuring spectral line profiles with high spectral and spatial resolution, to investigate the conditions in the solar atmosphere in three dimensions. Asymmetric line profiles provide information about gradients of physical quantities with height in the solar atmosphere, e.g. near-surface convection, acoustic waves, or magnetic reconnection. Moreover, due to the solar activity, the conditions can vary at a temporal scale of seconds to hours and also depend on the heliocentric {angle} on the solar disk.

The investigation of motions and other physical conditions in the solar atmosphere thus requires the precise measurement of the spectral line profiles and positions. A high temporal cadence, a well-defined averaging area, and a distinct repeatability for the measurements are imperative besides the high spectral resolution. With existing grating spectrographs and filter spectrometers, the typical wavelength accuracy amounts to around 2\,m\AA\ (${\rm\sim100\,m\,s^{-1}}$) in the visible range. Former techniques to calibrate the wavelengths of the solar spectrum are the usage of 
iodine cells \citep{Beckers1977}, spectral lamps \citep{2007A&A...468.1115L}, Fabry-P\'erot interferometer \citep{2014A&A...569A..77R}, or telluric lines as reference. However, the application of all given methods is limited, either by the low intensity level of the reference lines, their irregular distribution across the spectral range, or their known accuracy. The Laser Absolute Reference Spectrograph (LARS) overcomes all of these issues by using a laser frequency comb (LFC) to calibrate the solar spectrum recorded with a high-resolution echelle spectrograph, thus accomplishing a consistent accuracy {better by two orders of magnitude}. For solar physics, this makes LARS a unique instrument. 

\begin{figure}[htbp]
\begin{center}
\includegraphics[width=0.55\columnwidth]{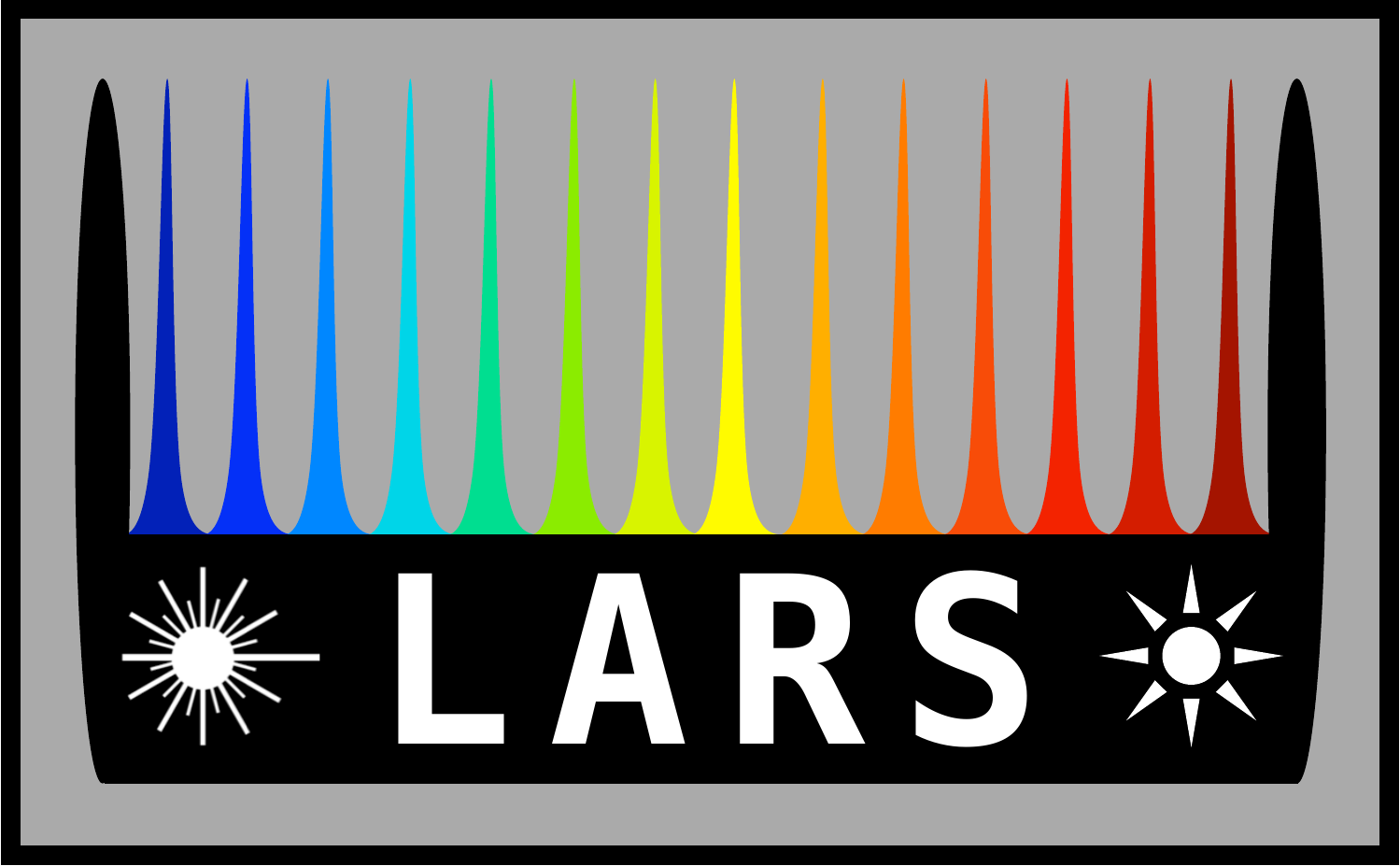}
\caption{Logo of the LARS instrument. The emission lines of the laser resemble the shape of a comb. The colors from blue to red indicate the operating range of the instrument in the visible. The symbols advert the assembly of laser and solar spectroscopy.}
\label{f_lars_logo}
\end{center}
\end{figure}

In astrophysical spectroscopy, a high spectral accuracy is imperative to reliably measure small-scale velocities of the gas at the solar surface. All systematic effects (like orbital motions, rotations, and gravitational shifts) on the solar spectrum need to be known or measured with sufficient accuracy, to guarantee the best quality for a careful and consistent data calibration. In addition, for many astrophysical investigations a high short-term and long-term stability is essential. To investigate large-scale flows, or global properties, like the convective blueshift, long averaging times are needed to eliminate small-scale convective and oscillatory motions that act as unwanted \glqq solar noise\grqq. With a high-resolution telescope like the German Vacuum Tower Telescope \citep[{VTT},][]{1985VA.....28..519S} observing only a small field-of-view, the significant systematic measurement of, e.g., the center-to-limb variation of the convective blue-shift may take weeks to month. They therefore require sensitive and very stable instruments to avoid the detrimental influence of unknown drifts, caused, e.g. by changes in the index of refraction of the ambient air. At the VTT, these issues are solved with the Laser Absolute Reference Spectrograph (LARS{, see logo in Fig.\,\ref{f_lars_logo}}), a system based on an LFC as an absolute calibration source. LARS provides an accurate wavelength calibration of each measurement over a large continuous wavelength range, and it guarantees long-term consistency of spectroscopic observations over months and years. 

In a pioneering work, \citet{Steinmetz+etal2008} demonstrated the feasibility of calibrating an astronomical spectrograph with a laser frequency comb. The authors used a comb operated in the infrared range to successfully calibrate the echelle spectrograph of the VTT. Between 2010 and 2013, an LFC-based wavelength calibration system was developed for the VTT spectrograph in a cooperation between the Kiepenheuer Institute for Solar Physics, Freiburg, the Max Planck Institute of Quantum Optics, Garching, and Menlo Systems GmbH, Martinsried. Initially, the system was planned to cover the visible spectral range from 500\,nm to 600\,nm with a mode separation of about 5\,pm. Due to the high spectral resolution of the  VTT spectrograph, the chosen mode separation was considered to be narrow enough to have a sufficient number of comb modes as calibration lines everywhere in the spectrum. At the same time the separation was large enough to observe clearly distinguishable comb lines with the shape of the spectrograph's points-spread-function \citep{Doerr+etal2012a,Doerr2015,Probst+etal2015}.

This paper describes the LARS instrument in its current science-ready configuration that was reached after substantial upgrades and modifications in 2016 and 2017, compared to the prototype version described in earlier works \citep{Doerr+etal2012,Doerr+etal2012a,Doerr2015,Probst+etal2015}. In Section \ref{sec_setup}, we briefly touch the properties of the telescope and spectrograph, followed by a description of the final setup of the LFC that now allows for turn-key operation of the instrument. The final part (Section \ref{sec_fibercoupling}) deals with the new opto-mechanical interface between the telescope and the single-mode fiber feed with fields-of-view between 1\arcsec\ and 10\arcsec\ in diameter. In Section \ref{sec_data}, we demonstrate the scientific performance of the instrument. Section \ref{sec_results} provides an outlook to the observing programs planned with LARS and the scientific opportunities with the instrument. 

Throughout the paper we stick to common practice in astrophysics and use air wavelengths when referring to spectral lines or observed wavelengths. The difference to vacuum wavelength (which is linked to frequency $\nu$ and speed of light $c$ through the well-known relation $\lambda\cdot\nu=c$) is about 0.14\,nm in the range around 500\,nm. An equation derived by \citet{Edlen1953, Edlen1966} is used to convert vacuum to air wavelengths, or vice versa.

%##############################################################################################
\section{Instrumental setup}\label{sec_setup}
In this section, we describe the instrument in its final science-ready configuration which was reached after a substantial upgrade of the frequency comb in 2016, and of the optical setup of the solar light channel in 2017. 

%%%%%%%%%%%% 
\subsection{LARS system}\label{sec_LARS_components}
\begin{figure}[htbp]
\begin{center}
\includegraphics[width=\columnwidth]{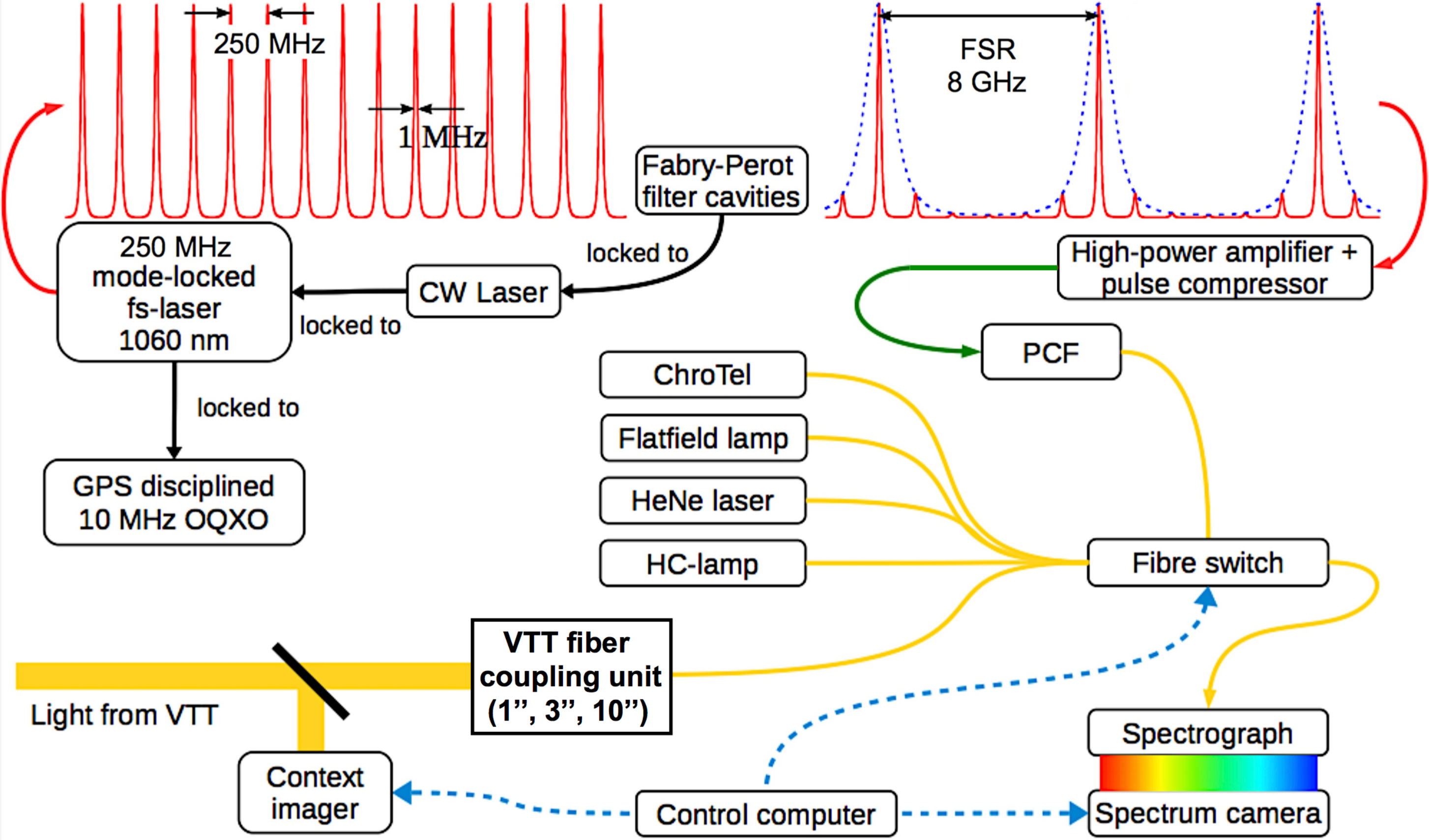}
\caption{LARS system. The femtosecond ytterbium laser operates at a wavelength of 1060\,nm with a repetition rate of 250\,MHz and is locked to a GPS-disciplined oscillator. The generated emission spectrum (upper left) passes a pair of identical Fabry-P\'erot cavities which filter the modes to an output repetition rate of 8 GHz (upper right). The frequency comb spectrum is amplified, and then broadened by a photonic crystal fiber (PCF). With single-mode fibers (yellow lines), the signal from each light source (LFC, sunlight, flatfield lamp, HeNe laser, hollow-cathode lamp, ChroTel) is guided to its own entrance port at the fiber switch. One output port is connected to the spectrograph (lower right). The light from the VTT falls onto a beamsplitter (lower left) and enters a fiber-coupling unit (1\arcsec, 3\arcsec, or 10\arcsec) for spectral measurements and a camera for context imaging of the surrounding solar region. {Acronyms are explained in Sects.\,\ref{sec_LARS_components} to \ref{sec_comb}.} Figure adapted from \citet{Doerr2015}.}
\label{f_components}
\end{center}
\end{figure}

LARS combines a number of optical subsystems to one powerful wavelength-calibrated solar spectrograph. A short overview of the full instrumental setup and the involved subsystems is given in this section. The schematic overview of the setup is shown in Fig.\,\ref{f_components}. The telescope and spectrograph are described in Sect.\,\ref{sec_telescope}. Within the scope of the instrumental upgrades, the laser frequency comb and the new fiber-coupling of the solar light will be explained in more detail in Sects.\,\ref{sec_comb} and \ref{sec_fibercoupling}.

LARS was developed to perform solar observations supported by an LFC as a source of frequency (or wavelength) calibration. The conjunction of both elements is illustrated in Fig.\,\ref{f_components}. The upper part sketches the generation of the frequency comb, the lower part shows the fiber-coupling of the solar light to the spectrograph. The sunlight collected by the VTT falls onto a cubic beam splitter which reflects $10\%$ of the incoming light to a camera imaging the solar region as it appears in a selected narrow wavelength range. The passing $90\%$ of the sunlight is guided to a fiber-coupling unit by which the light of a selected field aperture (1\arcsec, 3\arcsec, or 10\arcsec\ on the solar disk) is fed into a single-mode fiber. The integrated signal is then guided to a fiber switch which can rapidly change between eight entrance ports. The single output port is connected to the spectrograph for spectral observations. 

As depicted in Fig.\,\ref{f_components}, there are five different other light sources connected with fibers to the switch: i) A tungsten lamp which is used as a fiber-coupled flatfield lamp producing a continuous white-light spectrum for the calibration of the spectrograph camera; ii) a stabilized HeNe laser producing a sharp emission line to optimize the spectrograph alignment; iii) several hollow-cathode lamps which can be employed to measure the spectral emission of their atomic transitions; iv) the integrated full-disk light from the Chromospheric Telescope \citep[ChroTel,][]{2008SPIE.7014E..13K}; v) the laser frequency comb to absolutely calibrate the spectrograph and wavelength scale of all other spectra.

%%%%%%%%%%%% 
\subsection{Telescope and spectrograph}\label{sec_telescope}
For the sake of completeness, we briefly summarize the main properties of the VTT and its echelle spectrograph. Parts of the setup are sketched in the bottom of the setup overview in Fig\,\ref{f_components}. 
More information can be found on the VTT webpage\footnote{http://www.leibniz-kis.de/en/observatories/vtt/vtt-instrumentation/}. 

The 70\,cm telescope has a focal length of 45\,m and images a circular fraction of the solar disk (270\arcsec\ \O) at the main focal plane, with a plate scale of ${\rm 4.5\arcsec\,mm^{-1}}$. The entrance slit of the spectrograph is located at the focal plane in the direct, vertical beam. A 45$^\circ$ fold mirror allows to redirect the beam to various optical laboratories. For observations with LARS, the entrance slit is replaced by a fiber-coupling unit. A single-mode fiber (SMF) with a mode-field diameter of a few $\mu$m feeds the light (solar or artificial) to the spectrograph. This guarantees an identical illumination of the spectrograph for all light sources. The spectrograph design matches the telescope's ratio between focal length and aperture of 64, whereas the numerical aperture of the SMF corresponds to an f-ratio of 3.8. To adapt the SMF to the spectrograph, an asphere with a short focal length is mounted between the fiber and the entrance focal plane, to generate a beam with an f-ratio of about 64 that illuminates the large 79-grooves-per-inch echelle grating. Note that the spectrograph is not a classical, cross-dispersed echelle instrument. Instead, a pre-disperser produces a spectrum with low dispersion (about ${\rm1.6\,nm\,mm^{-1}}$) at the entrance focal plane of the main disperser. The main part of the spectrograph is an asymmetric all-reflective Czerny-Turner instrument with a 15\,m collimator and a 7.5\,m imaging mirror. This leads to a demagnification of the spatial scale by a factor of 2 compared to the telescope and to a dispersion of ${\rm19\,pm\,mm^{-1}}$ (${\rm \lambda=550\,nm}$). The spectral resolution R=$\Delta\lambda/\lambda$ amounts to about 800,000 at a wavelength of 550\,nm. To record the spectra, we use an ANDOR NEWTON CCD camera with a pixel size of 13.5\,$\mu$m and a chip size of 2048\,$\times$\,512 pixels. This results in a spectral field of view of 0.53\,nm at a wavelength of 550\,nm \citep{Doerr2015}.

%%%%%%%%%%%% 
\subsection{Key upgrade 1: The frequency comb}\label{sec_comb}
The major step from the expert-user prototype version of LARS to a turn-key systems was taken in May 2016.
The upgrade of the LFC now enables a user-friendly long-term stable operation of the instrument. The setup and its key upgrade are described in the following.

\begin{figure}[htbp]
\begin{center}
 {\includegraphics[trim=0cm 1cm 0cm 1cm,clip,width=\columnwidth]{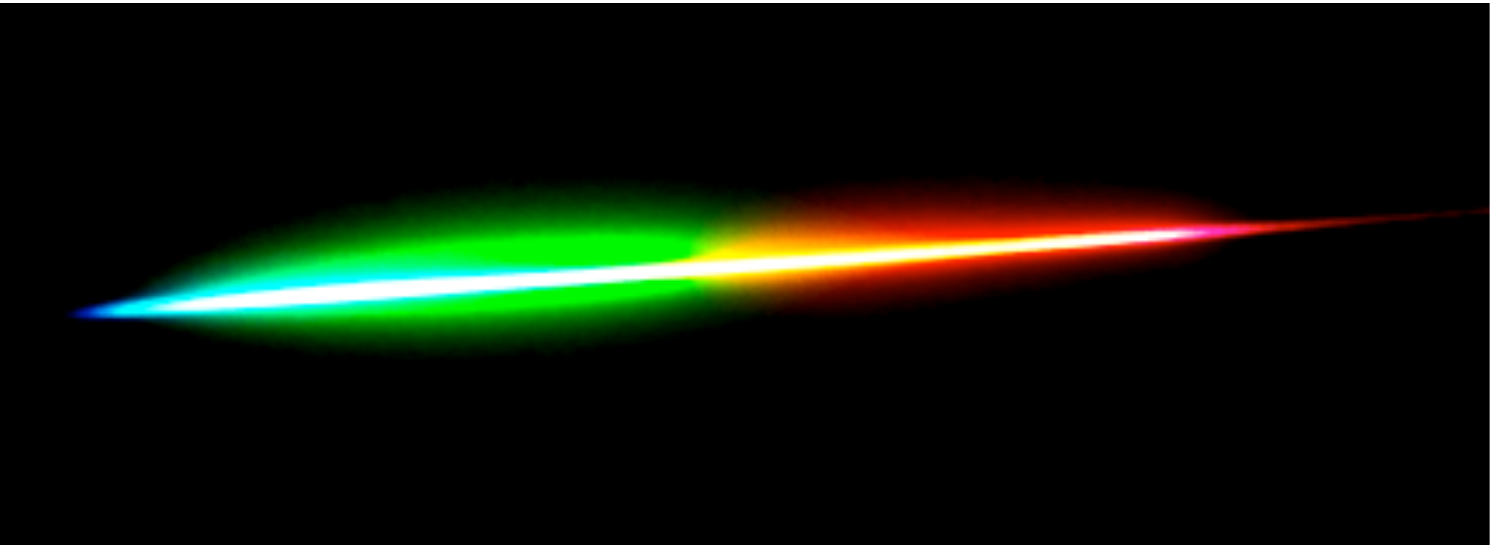}}
     \caption{Broadened spectrum of LARS. For illustration, the light leaving the optical fiber was dispersed with a small diffraction grating and projected to a nearby wall. Toward the red, the wavelength range is limited by the transmission range of the single-mode fiber (and by the sensitivity of the camera used for the photograph). The blue end depends on the optical power used for spectral broadening of the frequency comb. In this example we reach a wavelength coverage of about 480\,nm to 700\,nm.}
\label{f_colorspectrum}
\end{center}
\end{figure}

A femtosecond laser operating at a center wavelength of 1060\,nm generates a comb of equally spaced modes with a mode separation of 250 MHz. The native comb spectrum is sketched in the upper left of Fig.\,\ref{f_components}. The laser itself is referenced to a commercial 10\,MHz oven-controlled quartz crystal oscillator (OQXO), that is GPS-disciplined. To convert a frequency comb to an instrument suitable for astrophysical applications (\glqq astro-comb\grqq), the repetition rate and the spectral band of the laser have to be adapted to the capabilities of the spectrograph under consideration. This is achieved by sending the laser light through a pair of stabilized Fabry-P\'erot cavities, which transmit only every 32nd comb mode. The cavities are locked to the repetition rate of the LFC by the transmission signal of a continuous wave (CW) laser. The resulting output signal is sketched in the top right corner of Fig.\,\ref{f_components}. The output repetition rate (mode separation) amounts to 8.0\,GHz. The signal is amplified in a cladding-pumped Yb-doped fiber amplifier, and thereafter broadened in a tapered photonic crystal fiber (PCF). The final, spectrally broadened LFC signal is shown in Fig.\,\ref{f_colorspectrum} with low dispersion. The light exiting the PCF is guided through a single-mode fiber to an additional free-space unit with a knife-edge (not depicted in Fig.\,\ref{f_components}). By blocking parts of the light beam, the intensity level of the comb light is adjusted to the continuum level of the solar spectrum. The regulated comb signal is coupled to another single-mode fiber and guided to the fiber switch. The general design of an astro-comb was already described by \citet{Steinmetz+etal2008} and \citet{Wilken+etal2012}. \citet{Doerr2015} and \citet{Probst+etal2015} discussed the prototype version of the present instrument in great detail. 

\begin{figure}[htbp]
\begin{center}
\includegraphics[width=\columnwidth]{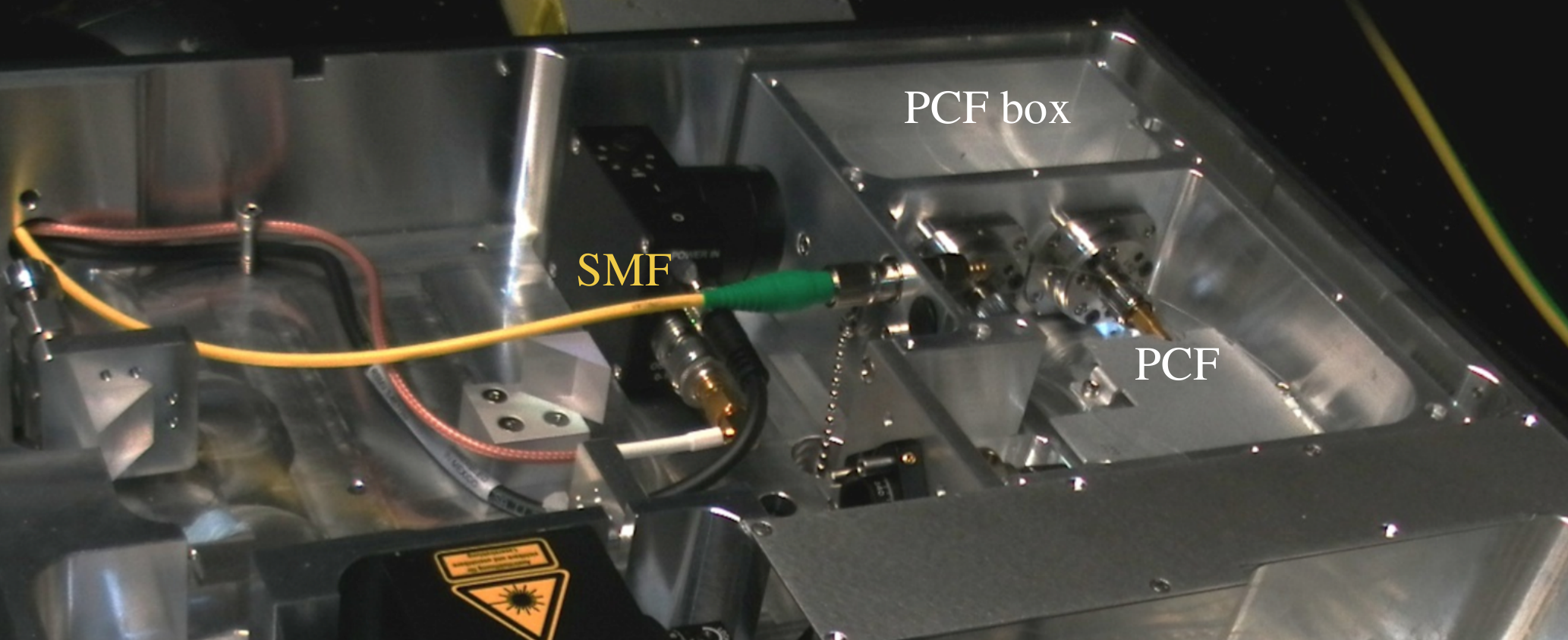}
\caption{The photonic crystal fiber (PCF) is connectorized at both ends, with sealed end facets, and is mounted in a segregated housing (box in the right part of the picture) for quick and easy replacement. The PCF output is coupled to a single-mode fiber (yellow) and guided to the knife-edge unit and fiber switch.}
\label{f_PCFbox}
\end{center}
\end{figure}

%Schematic view: A solid elliptical core (bright inner cell of the cross-section) is surrounded by air cells (black). To broaden the comb spectrum, the fiber is tapered to a core diameter of 536\,nm. Figure adapted from \citet{ediss18701} showing the cross-section of the previous fiber.

In the prototype version \citep{Doerr2015}, the infrared laser signal was translated into the visible range by using a second harmonic generator, which doubled the frequency. An earlier version of the tapered PCF then broadened the laser signal to a range from 460\,nm to 700\,nm. In the meantime, the PCF technique matured by optimizing the taper geometry\footnote{The tapered elliptical core of the PCF spans only few micrometers.} and PCF structure, now broadening the fundamental comb spectrum to a substantially wider range, making the use of a second-harmonic generator obsolete. The frequency comb spectrum now spans from 480\,nm to 1300\,nm. This was a major step forward simplifying the optical setup and, at the same time, substantially increasing the operational stability. The new PCFs do not only provide much larger spectral broadening, they also have a significantly longer life time of several thousands of operating hours. They are readily available as plug-and-replace units which can be exchanged quickly. The PCF and its box are shown {in Fig.\,\ref{f_PCFbox}}. As a by-product of this development, all optics are now fully contained in optical enclosures, which again improved the stability and decreased the sensitivity to changes in the ambient conditions. The new software package controls all instruments para\-meters and regulates them automatically to the optimized setting. Eight weeks of observation time in 2016 have verified an uninterrupted system stability of up to several days. 

In the prototype version, the native laser mode separation of 247.5\,MHz had been filtered to 5.445\,GHz. This number translated into a mode separation of 4.5\,pm (at $\lambda=500$\,nm), well resolved by the spectrograph. On the other hand, the mode spacing was large enough to clearly identify the wavelength of an individual laser mode due to its proximity to a solar spectral line. Once the wavelength (and thus the frequency) of a single laser mode is identified, all other modes are exactly known, due to the given offset frequency and the constant spacing of the frequency modes. For the final setup we decided to change the mode separation from 5.445\,GHz to 8.0\,GHz. Thus, only every 32nd mode is transmitted by the Fabry-P\'erot cavities. The mode separation of 8.0\,GHz (6.67\,pm at $\lambda=500$\,nm) is still convenient for the visible wavelength range and for the VTT spectrograph with its extremely high spectral resolution. In addition, the larger repetition rate would enable the use of LARS also for the near infrared GRIS instrument \citep{Collados+etal2012} at the GREGOR solar telescope \citep{Schmidt+etal2012}. 

{Intrinsically, the LFC provides a  wavelength accuracy at level of a few ${\rm cm\,s^{-1}}$. The stability of the GPS-disciplined reference oscillator ultimately limits the frequency accuracy of the comb to an absolute value of only ${\rm 3\,mm\,s^{-1}}$. It is the spectral filtering of the native comb with the Fabry-P\'erot cavities which introduces a minimal mismatch between the FPC transmission peaks and the equidistant comb peaks. As a result, the side-modes are suppressed with a slightly asymmetric weight. Then, fitting the transmitted modes with a Gaussian leads to a worst-case shift of ${\rm 2.1\,cm\,s^{-1}}$. In addition, during the power amplification for spectral broadening, the side-modes are unavoidable re-amplified to an intensity of around $-67$dB relative to the transmitted mode. This imperfection is tremendously small compared to other effects \citep{Wilken+etal2012}, but enters an additional defective shift of the LFC line centroids of 5\,kHz, or ${\rm 3\,mm\,s^{-1}}$ at 630\,nm \citep{Doerr2015}.}

%%%%%%%%%%%% 
\subsection{Key upgrade 2: Fiber-coupling of the solar light to the spectrograph}\label{sec_fibercoupling}
The second major upgrade of LARS was performed in April 2017. The optical setup of the solar channel was restructured to facilitate a quick change between three available field apertures for solar observations. On this occasion, the majority of the optical components was replaced.

\begin{figure}[htbp]
\begin{center}
\includegraphics[trim=0cm 0cm 0.8cm 0.5cm,clip,width=\columnwidth]{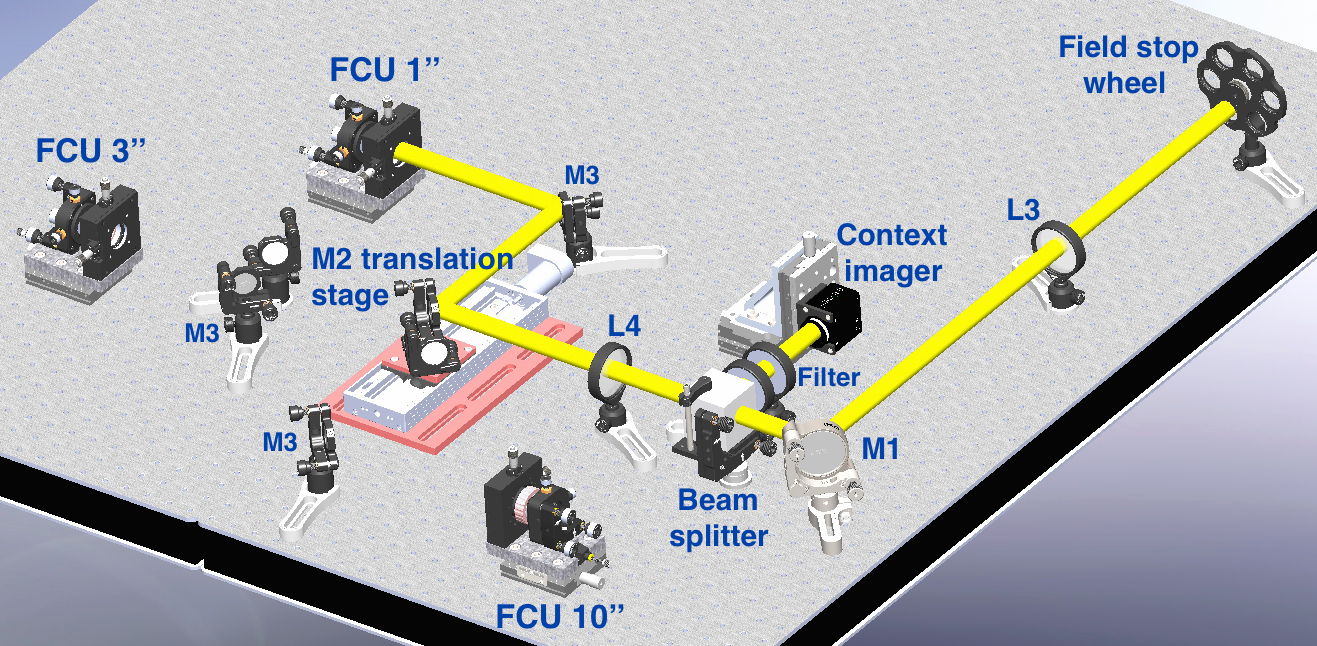}
\caption{New optical setup for solar observations with LARS. The fiber-coupling units (FCU), lenses (L), mirrors (M), translation stage, and Context Imager are sketched.}
\label{f_optics}
\end{center}
\end{figure}

As illustrated in Fig.\,\ref{f_components}, LARS observes the solar light simultaneously with two cameras. The Context Imager records a two-dimensional image of the solar region captured by the telescope. At the same time, a small circular field in the center of that region is fed into a single-mode fiber and its field-integrated spectrum is observed by the spectrograph camera. Three fiber-coupling units are available to measure the sunlight with an aperture corresponding to either 1\arcsec, 3\arcsec, or 10\arcsec\ on the sky. {For solar observations, such a distinction of the integrated region is necessary to perform exclusive and qualitative measurements of the atmospheric conditions for different spatially-resolved solar features, e.g., solar granules with a size of 1\arcsec or umbrae with a diameter of 10\arcsec.} A pinhole in the final focus limits the field, a collimating lens with short focal length feeds the light to the fiber facet\footnote{Since it is not possible to image the observed aperture on the tiny fibre core, we project the telescope pupil on the fibre facet. This ensures that information from the whole aperture is coupled to the fibre. But owing to the Gaussian acceptance profile of the fibre, the inner regions of the aperture are covered with higher efficiency than the outer regions. {Nevertheless, the optics were designed such that the coupling efficiency in the outer regions of the aperture does not drop below 60\%.}}. For observations, a change of the fiber-coupled aperture therefore involved a manual exchange of the pinhole and lens in front of the fiber. Since this change included a sensitive optical alignment to maximize the coupling efficiency, this process was laborious and time-consuming. Therefore, the basic idea of the upgrade was to install three individual pre-aligned fiber-coupling units, one for each field aperture. A motorized translation stage with two mounted silver-coated mirrors guides the light to one of the fiber-coupling units. This automatization now allows a selection of the different apertures within seconds. 

The old setup \citep{Doerr2015} consisted of a X95 rail and carrier system to align the optical components with the light path. As part of the upgrade, the rail system was replaced by two 60\,mm optical breadboards for the new optics. The inner honeycomb structure optimizes the damping of vibrations and acoustic disturbances below 100\,Hz. The new optical design is displayed in Fig.\,\ref{f_optics}. The transfer optics consisting of two lenses (L1 and L2, not shown here) with focal lengths of 1500\,mm relay the science focus of the VTT toward a stop wheel with a set of removable blends. Besides the free field, three pinholes with different field apertures (1\arcsec, 3\arcsec, 10\arcsec\ on the sky) are mounted at the wheel to align the light path to the fiber and context camera. A third 50\,mm diameter achromatic lens (L3) with a focal length of 300\,mm collimates the light to the silver-coated mirror (M1).  The high-precision two-axes adjusters of the mirror mount facilitate the optimized alignment of the light beam. After the mirror, a cubic beamsplitter reflects 10\% of the incoming light to the Context Imager. A combination of a 50\,mm lens and an interference filter focus the wavelength-filtered light onto the CCD chip of the Context Imager camera. Information about the data acquisition is given in Sect.\,\ref{sec_context_imager}.

\begin{figure}[htbp]
\begin{center}
\includegraphics[width=0.9\columnwidth]{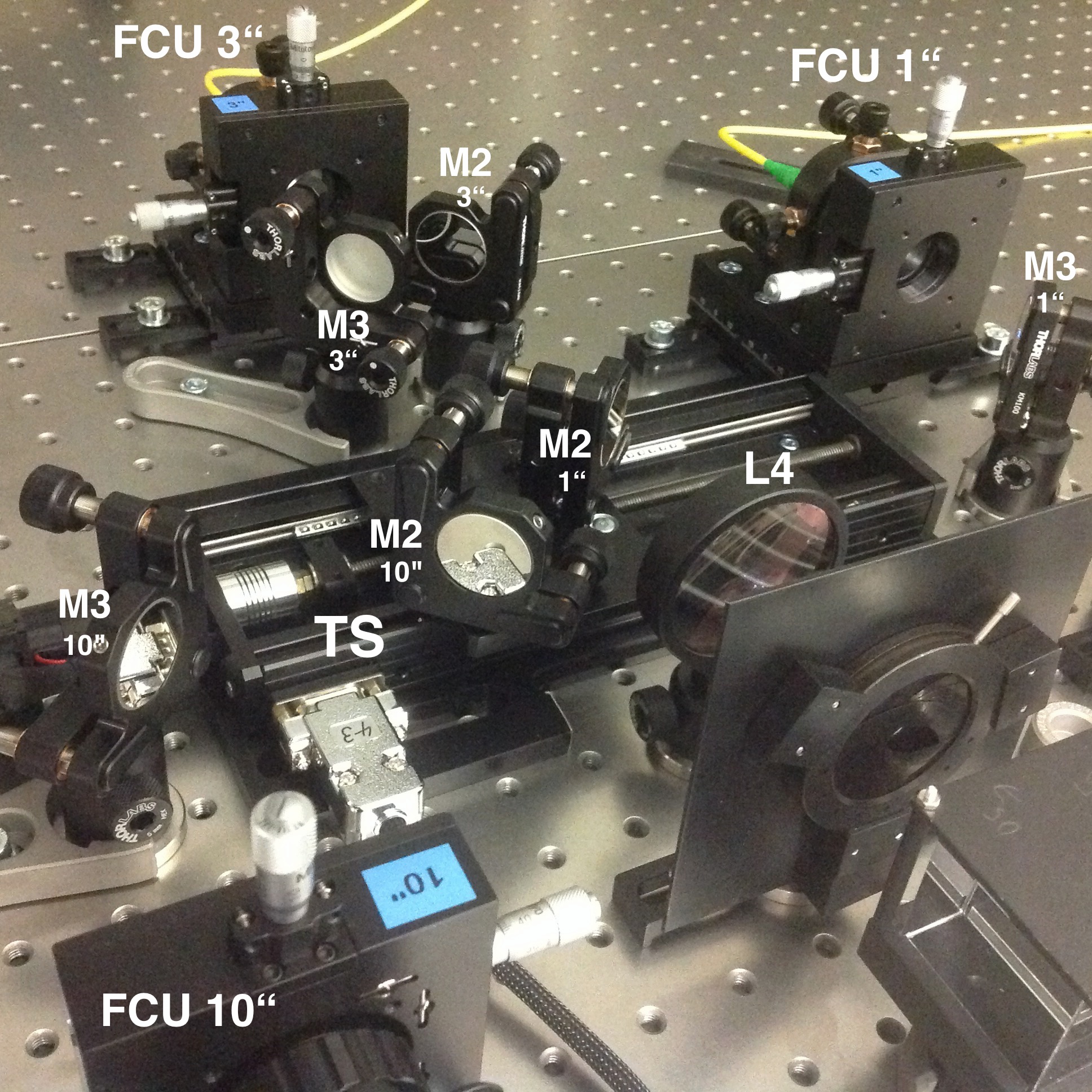}
\caption{Fiber-coupling units. The solar light of the respective field-of-view (1\arcsec, 3\arcsec, or 10\arcsec) is fed to the fibers. The integrated signal is guided to the spectrograph. Compare Fig.\,\ref{f_optics}.}
\label{f_fiber_coupling}
\end{center}
\end{figure}

The remaining 90\% of the light passes the beamsplitter toward the fiber-coupling units (see Fig.\,\ref{f_fiber_coupling}). A 50\,mm achromatic lens (L4) with a focal length of 300\,mm focuses the light toward the pinhole of the fiber-coupling unit. The heart of the system upgrade is the increase from one to three individual fiber-coupling units. A motorized translation stage (TS) manufactured at the Kiepenheuer-Institut handles the beam guiding. Two silver mirrors (M2) with diameters of 25.4\,mm are mounted on the moveable stage with an angle of incidence of $\pm45^{\circ}$ to the incoming light beam. Three pre-adjusted positions (first, second, or no mirror) can be selected with micrometer repeatability. In combination with an additional folding mirror (M3), an identical length of the light path to all fiber-coupling units is guaranteed. All mirrors are mounted in two-axis adjusters to enable a precise and stable alignment.

\begin{figure}[htbp]
\begin{center}
\includegraphics[trim=0cm 10.5cm 0cm 4.5cm,clip,width=0.75\columnwidth]{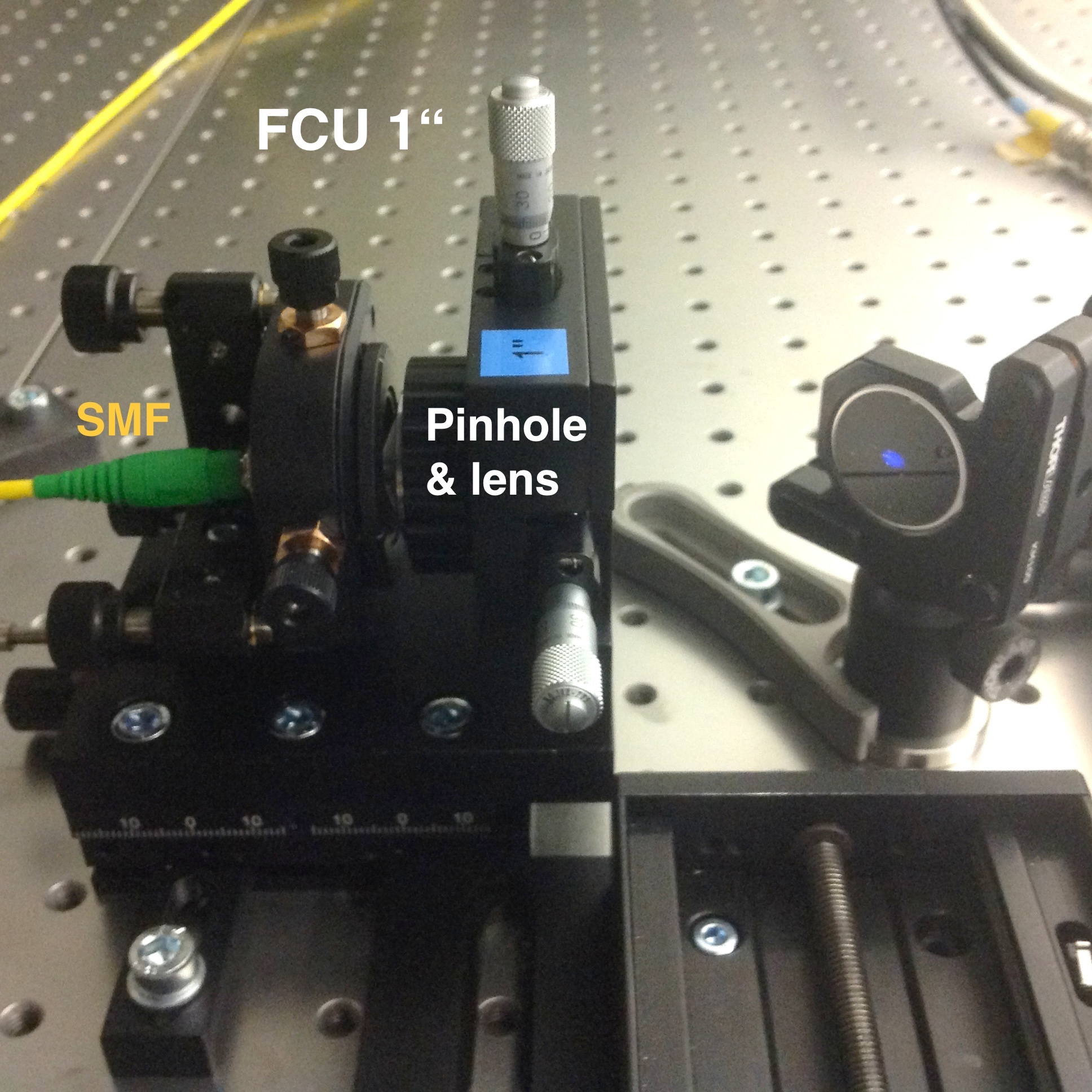}
\caption{Fiber-coupling unit with pinhole, lens and single-mode fiber (SMF) of the 1\arcsec\ field.}
\label{f_fiber_1arcsec}
\end{center}
\end{figure}

The fiber-coupling unit feeds the sunlight into the optical single-mode fiber. The optical components are displayed in Fig.\,\ref{f_fiber_1arcsec}. A circular pinhole is drilled into a brass plate to limit the field aperture to 1\arcsec, 3\arcsec, or 10\arcsec\ on the solar disk. The lens with short focal length collimates the residual light beam toward the fiber facet. To maximize the coupling efficiency into the fiber, the fiber-coupling unit in total features nine degrees of freedom for orientation and alignment. The whole unit is placed on top of a dovetail translation stage which can shift the unit by 25\,mm in the direction of the light path. The pinhole and lens can be moved in two axes, vertically and horizontally with respect to the fiber. The fiber mount itself can be adjusted along six axes. To bring the fiber into the optimal position along the direction of the light beam, the unit is placed on top of a one-axis translator. The fiber head itself has a five-axes mount for optical alignment in X, Y, and Z direction, as well as tip and tilt. The light coupled to the fiber (yellow cable in Fig.\,\ref{f_fiber_1arcsec}) is guided internally to the fiber switch. When the VTT channel is selected by the control software, the light enters the output fiber which is connected to the spectrograph. 

All optical fibers of LARS are single-mode fibers. The fibers are specified for a peak transmission in the range from 450\,nm up to 700\,nm. Due to the operating range of the frequency comb starting at 480\,nm, this limits LARS observations to the visible part of the spectrum from 480\,nm to 700\,nm. Toward the red, the damping due to the fiber necessitates exposure times of a few seconds to perform spectral observations. Unlike multi-mode fibers, single-mode fibers guide only one propagation mode. This brings the advantage of well-defined beam properties which are completely independent of the input coupling parameters. A stable and uniform illumination of the spectrograph grating is achieved, which in turn enables the high calibration accuracy of the spectra. On the other hand, the coupling efficiency of the incoming light is very low. But since the Sun (compared to other stars) is an extended source with high intensity, the light level transmitted through the single-mode fiber is always adequate to allow for high-cadence ($\sim1\,{\rm Hz}$) solar spectroscopy.

%%%%%%%%%%%% 
\subsection{Instrument control}
\begin{figure}[htbp]
\includegraphics[width=\columnwidth]{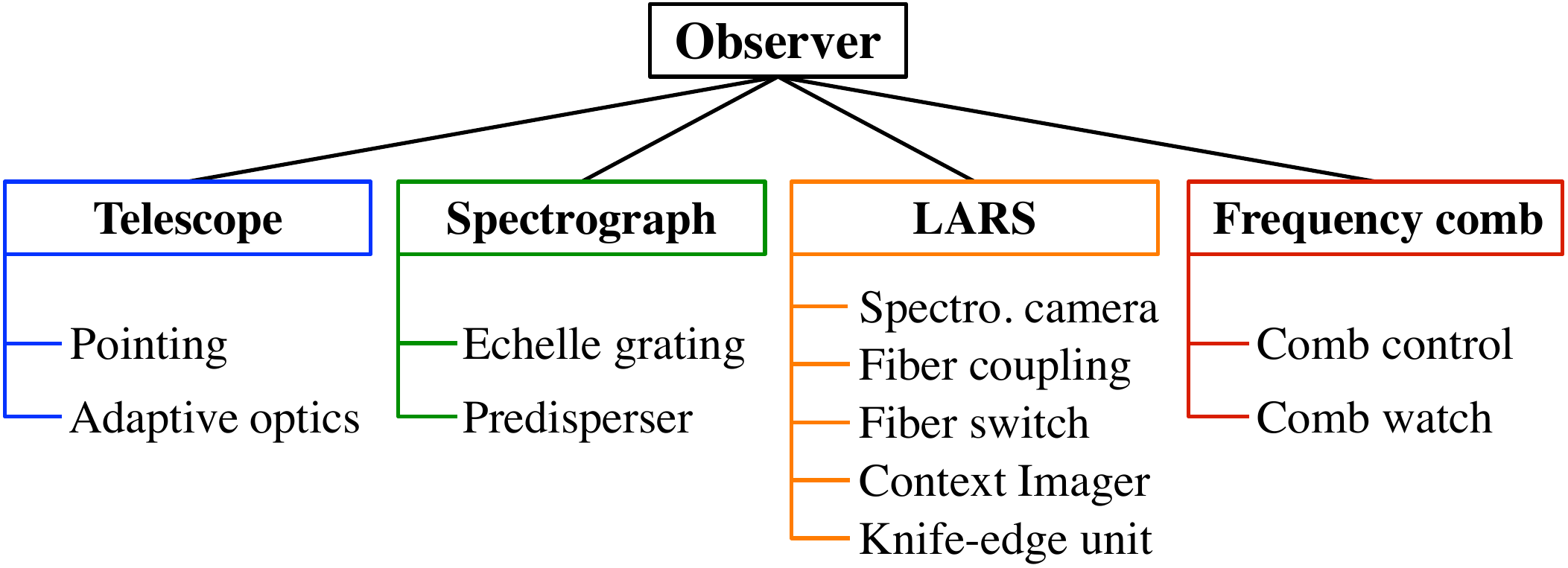}
\caption{Schematic overview of the instrument control. The observer operates four individual control units and their sub-units.}
\label{f_instrument_control}
\end{figure}

For spectroscopic observations with LARS, a number of independent control units are involved. A schematic overview is given in Fig.\,\ref{f_instrument_control}. It contains: (i) the telescope operation system and pointing, as well as the adaptive optics (AO) system, (ii) the spectrograph with its echelle grating and predisperser, (iii) the different units operated by the LARS control software, these include the spectrograph CCD camera, the translation stage selecting the active fiber-coupling unit, the fiber switch to select the input source for the spectrograph, the Context Imager, and a knife-edge unit to regulate the light level of the frequency comb, and (iv) the frequency comb itself with its operation control and monitoring. The four main units are operated independently by the observer. The following sections describe the last two units. 

\subsubsection{LARS control}
The LARS control software commands the five sub-units listed in Fig.\,\ref{f_instrument_control} (third column from the left). It sets the observational parameters for the spectrograph camera, like the integration time of the CCD, the read-out area and binning mode, the temporal cadence and number of repetitions within the sequence, and the camera cooling. It also controls the fiber switch to alternate between the selected input channels of the different light sources. The Context Imager can be operated individually or triggered by the spectrograph camera. The knife-edge unit is driven to regulate the light level of the frequency comb. The control software package {was supplemented by} the automated translation of the fiber-coupling units for the field-of-view. Before the optical upgrade described in Sect.\,\ref{sec_fibercoupling}, the change of the fiber-coupling unit had to be performed by hand, including optical alignment to maximize the coupling efficiency. This was laborious and time-consuming. The new motorized translation stage and pre-aligned fiber-coupling units allow an automated selection of the different fields-of-view within a few seconds with a high repeatability. Its operation is now included in the LARS control software as part of the upgrade.

\subsubsection{Frequency comb control}
Although being one of several light sources connected to the fiber switch, the LFC and its control system are operated  independently from the rest of the instrument (see right column in Fig.\,\ref{f_instrument_control}). The new version of the comb control system that came with the upgrade in 2016 (see Sect.\,\ref{sec_comb}) now allows for an automated operation of the comb. This was the decisive step to convert the whole system from a complex prototype into a (still complex, but easily manageable) turn-key instrument. The LFC control system can be operated remotely, e.g., from the VTT control room, or from any authorized computer outside the observatory. This remote capability allows for remote support by experts in case of a malfunction. To check or reinspect the correct operation of the LFC system, an internal log file with all context information is written.

\begin{figure}[htbp]
\begin{center}
\includegraphics[trim=1.0cm 0.7cm 0cm 1.2cm,clip,width=0.92\columnwidth]{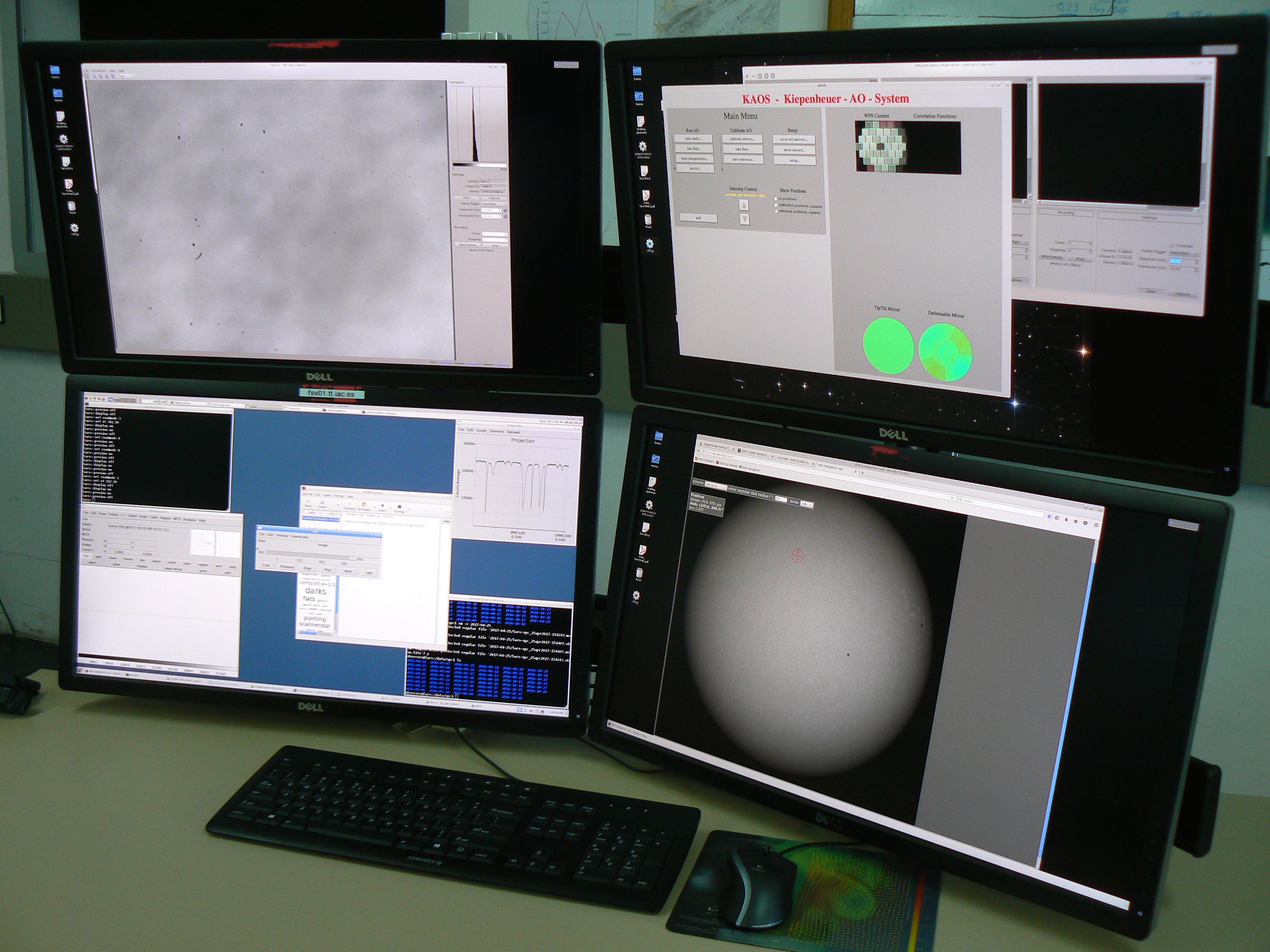}
\caption{Remote control station at the VTT. All instruments and devices needed for LARS observations are accessible via remote desktop tools and are displayed on an array of screens.}
\label{f_remote}
\end{center}
\end{figure}

\subsubsection{Remote Control Station}
To facilitate solar observations with LARS, all required control screens can be bundled as remote sessions at the remote control station (see Fig. \ref{f_remote}) in the control room of the VTT. Experienced users are thus able to perform LARS observations single-handedly. All remote sessions can also be controlled from any other authorized computer.

%##############################################################################################
\section{Observations and data processing}\label{sec_data}
In this section, we discuss the observations which are performed with the two cameras of LARS -- the Context Imager (Sect.\,\ref{sec_context_imager}) and the spectrograph camera (Sect.\,\ref{sec_spectrograph_camera}). We thereby focus on the acquired data and give a summary of the data calibration for both channels.

\subsection{Context Imager}\label{sec_context_imager}

\begin{figure}[htbp]
\begin{center}
\includegraphics[width=\columnwidth]{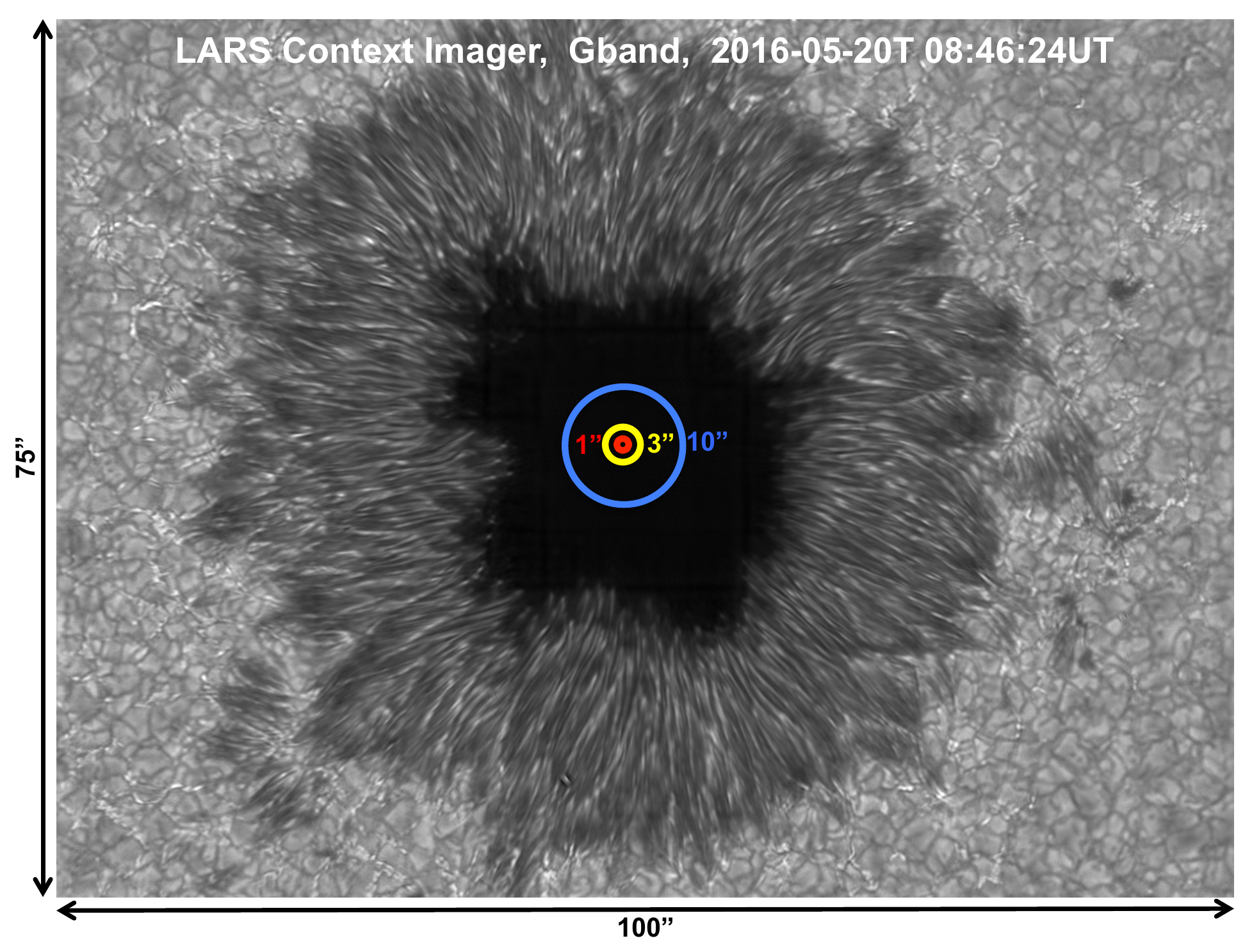}
\caption{Sunspot observed with the Context Imager of LARS. The image sequence was recorded on May 20th 2016 at 08:46\,UT with an interference filter centered on the spectral G-band around 430\,nm. The Speckle-code KISIP was used to reconstruct the displayed image. The full field-of-view covers a size of $100\,\arcsec\times75\,\arcsec$. The light from a circular region with a diameter of either 1\arcsec\ (red), 3\arcsec\ (yellow), or 10\arcsec\ (blue) is integrated to a fiber and guided to the spectrograph.}
\label{f_sunspot_fov}
\end{center}
\end{figure}

\paragraph{Observation:} The Context Imager records a two-dimensional image of the Sun. The sunlight is filtered to one narrow wavelength range of typically 1\,nm or less. Various spectral pre-filters can be inserted to yield context information, e.g. about photospheric, chromospheric, or magnetic dynamics in the solar atmosphere. An example of a recorded context image is displayed in Fig.\,\ref{f_sunspot_fov}. 
It shows a fully-developed sunspot recorded on May 20th 2016 at 08:46\,UT close to the solar disk center at a wavelength of 430.8\,nm (G-band). The reasons to select this spectral band were the simplified distinction between magnetic and non-magnetic regions, and the naturally higher spatial resolution in the blue. With the appropriate pre-filter, the Context Imager can be operated at any wavelength in the visible. The CCD camera has a detector size of $1360\times1024$\,pixel, translated into a full field-of-view of $100\arcsec\times75\arcsec$ on the Sun. The Context Imager can be synchronized to the spectrograph camera. Then, for every acquired spectrum, a context image is recorded simultaneously. The Context Imager can also be operated independently. Sequences with up to 20 frames per second and exposure times of a few milliseconds can be achieved. {Adaptive optics can be enabled to correct for image distortions caused by atmospheric turbulences. Image reconstruction techniques are then applied to reach a spatial resolution close to the diffraction limit of the telescope.} The two-dimensional image serves as context information for the spectroscopic observation. The three different fields-of-view and their alignment with respect to the context image are indicated in Fig.\,\ref{f_sunspot_fov} as red (1\arcsec), yellow (3\arcsec), and blue (10\arcsec) circles. 

\paragraph{Data calibration:} The raw data recorded with the Context Imager contain several defects like dissimilar pixel sensitivities, dust or dirt on the detector, and atmospheric distortions of the solar image which are corrected by the data calibration. The set of calibration data includes a sequence of \glqq dark\grqq\ and \glqq flatfield\grqq\,\footnote{The telescope performs a fast movement over a Quiet Sun region close to the observed solar target while the camera takes an image sequence long enough to smear out all solar structures.}\ images to evaluate the background signal and intensity defects, and to calculate the average gaintable image for the flatfield-correction of the image sequence. 
\begin{figure}[htbp]
\includegraphics[trim=1.2cm 5.4cm 3.05cm 2.18cm,clip,width=\columnwidth]{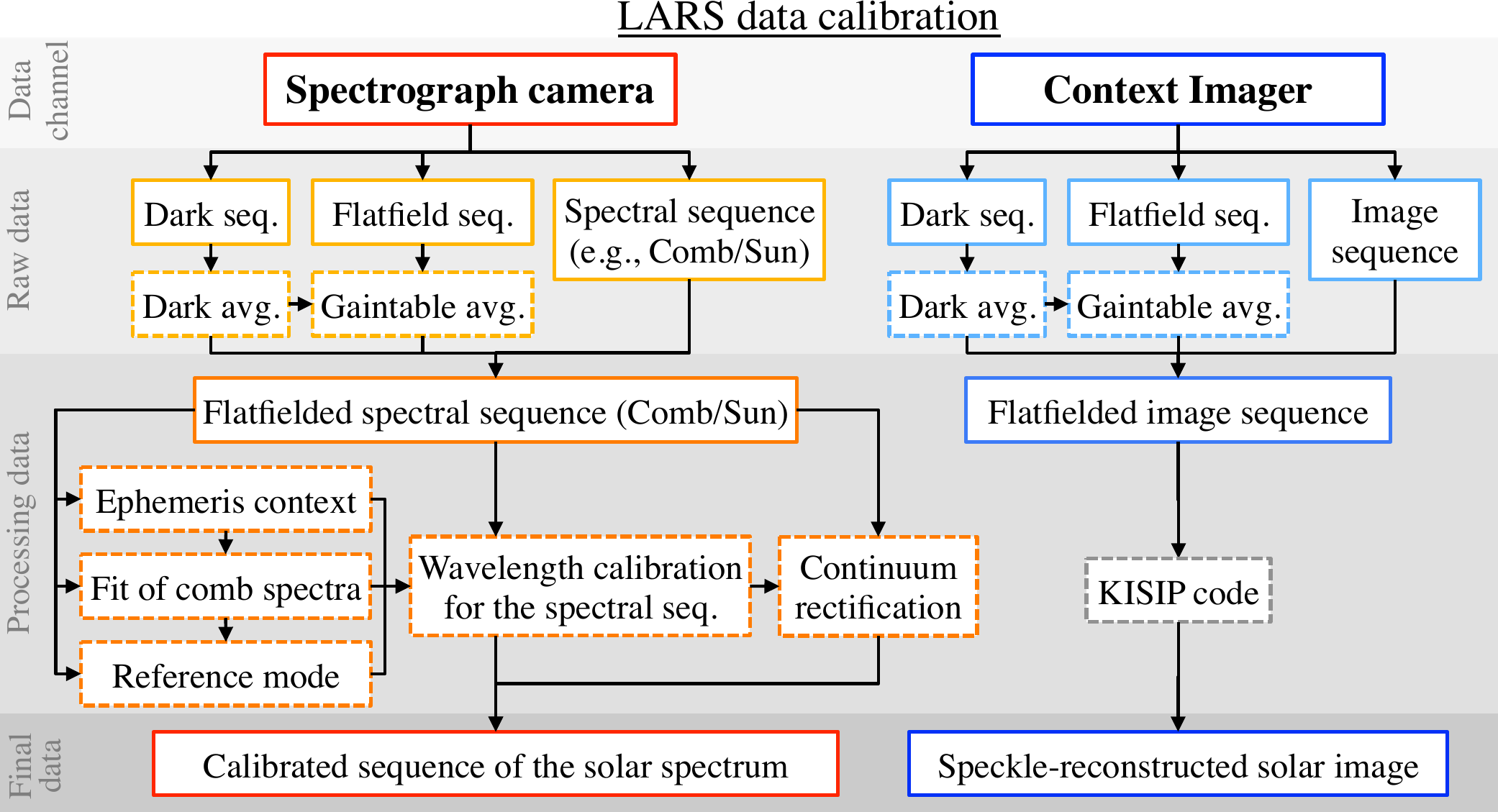}
\caption{Calibration scheme for LARS data. The data reduction from the raw to the final state is displayed for both LARS channels -- the spectrograph camera (red colors) and the Context Imager (blue colors). Black arrows indicate the direction of processing. Raw, processing, and final state products are drawn with solid borders around the text box. Intermediate calibration steps and tools are indicated by dashed borders.}
\label{f_data_calibration}
\end{figure}
The calibration scheme for the Context Imager data is shown on the right side of Fig.\,\ref{f_data_calibration}. As a matter of routine, a set of context images\footnote{An AO-stabilized image of an US Air Force test target is taken to gain information on the spatial resolution. A pinhole image is obtained to verify the position of the AO lock point. Finally, context information on the location of the fiber and the size of the integrated region is taken for the spectral observation. The respective field stop at the filter wheel is inserted and a field-of-view image of the fiber is recorded.} is recorded which are not named in the processing scheme. The flatfielded image sequences can already be used for scientific investigations but still suffers from residual atmospheric distortions which can not be coped by the adaptive optics system. Sequences recorded with a rate of 10--20 frames per second and a short exposure time ($<50$\,ms) are post-facto reconstructed employing the Kiepenheuer-Institute Speckle Interferometry Package \citep[KISIP,][]{2008SPIE.7019E..1EW} to approach the a diffraction-limited spatial resolution.

\subsection{Spectrograph camera}\label{sec_spectrograph_camera}

\begin{figure}[htbp]
\begin{center}
\includegraphics[width=\columnwidth]{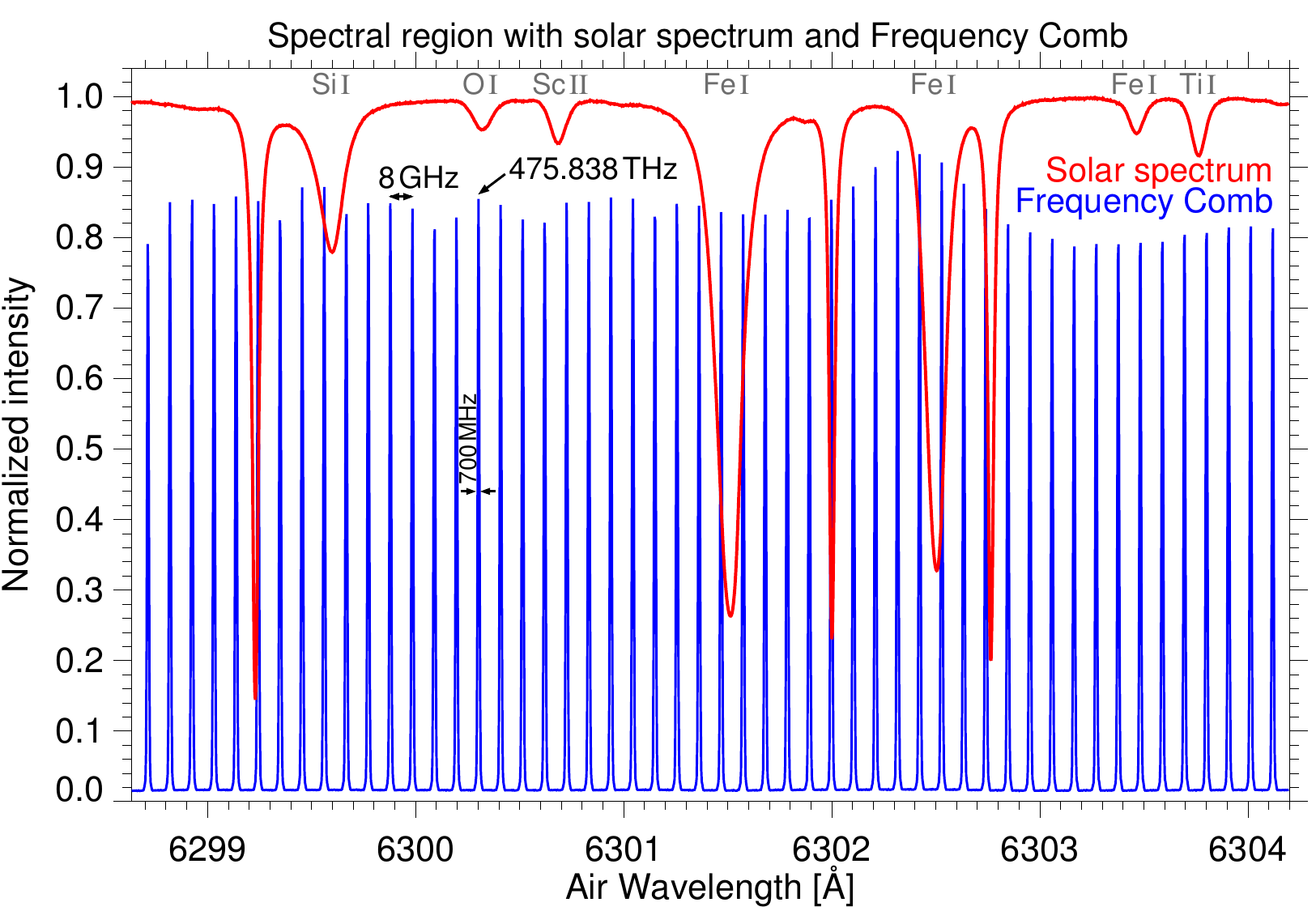}
\caption{Spectral region with a typical cycle of solar spectrum and frequency comb signal. The spectrograph camera records a spectral region with a wavelength width of 5.6\,\AA, here centered around 6301.5\,\AA. The solar spectrum in red was normalized to the continuum intensity and consists of several photospheric lines (elements in gray). The three unnamed narrow lines are telluric O$_2$ lines.
The frequency comb spectrum is overlaid in blue and consists of 52 equally spaced (8\,GHz) emission modes with a full width at half maximum of 700\,MHz. One frequency mode (here 475.838\,THz) has to be unambiguously identified through its proximity to a spectral line core. Note that the solar and the comb lines are indeed observed sequentially, to avoid any deterioration of the solar line profile through the superposition of the comb lines.}
\label{lars_spectrum}
\end{center}
\end{figure}

\paragraph{Observation:} The spectrograph camera records a one-dimensional spectrum of the selected input source (see Sect.\,\ref{sec_LARS_components}). We use an ANDOR NEWTON CCD camera with a pixel size of 13.5\,$\mu$m and a chip size of 2048\,$\times$\,512\,pixels. Owing to the single fiber feed of the spectrograph, the signal is concentrated in only few (typically two to three) adjacent pixel rows of the CCD chip. {However, due to effects of instrument internal seeing, or changes of the ambient air pressure and temperature, the spectrum suffers from potential drifts. Detailed characterizations of the spectrograph stability \citep{Doerr2015} revealed that long-term variations of the environmental conditions can lead to shifts of the LFC spectrum of a few hundred meters per second within several hours. In addition, the short-term variation shows an instrumental jitter at the scale of 10\,s with an amplitude up to a few ${\rm m\,s^{-1}}$. With the LFC spectra recorded at a cadence of a second or below, we can account for these drifts. To consider spatial shifts of the spectrum across the camera sensor, a} region of around twenty rows around the illuminated pixels is read out and binned perpendicular to the dispersion axis. With a dispersion of ${\rm19\,pm\,mm^{-1}}$ at a wavelength of ${\rm \lambda=550\,nm}$, the 2048\,pixel wide spectrum covers a spectral range of 0.53\,nm. The spectral resolution of the spectrograph amounts to about 800,000 at that wavelength. A typical observation sequence with LARS consists of the repetition of a two-channel cycle. For solar observations, one cycle constitutes of the successive measurement of the frequency comb spectrum and the solar spectrum. An exemplary cycle is shown in Fig.\,\ref{lars_spectrum}. 
Typical integration times are around 1\,s, depending on the solar target. {In case of a Quiet Sun region at disk center and a single exposure for 1\,s, a signal-to-noise of 200 is reached for the spectral continuum around 550\,nm. At this signal level, dark current and read noise do not play a significant role. As reported in \citet{Doerr2015}, the signal level decreases only to a minor fraction when changing between the 10\arcsec, 3\arcsec, and 1\arcsec\ fiber-coupling units. This has the fundamental reason that the coupling efficiency of the multi-mode light source to a single-mode fiber increases with decreasing field-of-view, in our case by one magnitude from 10\arcsec\ to 3\arcsec, and 3\arcsec\ to 1\arcsec\ each.}

{Given the temporal scales of seconds, very fast camera readout is not important. During the cycle, the} fiber switch can change the input channel within 2\,ms. This sets a total cycle time of a few seconds. By repeating the cycle, the frequency comb and solar spectra are observed alternately. So the solar spectrum can be calibrated by the interpolation between the preceding and succeeding comb spectrum. The user defines the number of repetitions and, by this, the observation time of the sequence. {The fully-automated data capture software writes} the data sequence as one file in FITS format. The first dimension contains the 2048\,pixels, the second dimension successive cycles (or time). The metadata required by the data pipeline is written into the FITS header.

\paragraph{Data calibration:} The observed raw data of the spectrograph camera still contains systematic defects like dissimilar pixel sensitivities and dust on the detector, the background readout, and large-scale gradients in the spectrograph transmission. A set of \glqq dark\grqq\ and \glqq flatfield\grqq\ sequences is recorded to reduce these errors from the spectral sequences. The scheme of the spectral data calibration is shown in Fig.\,\ref{f_data_calibration} on the left side. Two short \glqq dark\grqq\ sequences taken with the covered camera define the readout and background signal of the camera. Since the camera can be cooled to $-90^\circ$C, the dark signal is very low. The exposure times of the darks have to be the same as for the spectral sequence (e.g., \glqq LFC/Sun\grqq) and \glqq flatfield\grqq\ sequence, respectively. A tungsten lamp emits the continuous flatfield spectrum which contains the instrumental errors when recorded by the camera. The exposure time for the flatfield is adjusted to match the continuum intensity level of the solar spectrum. The raw data (yellow boxes in Fig.\,\ref{f_data_calibration}) is entered to the semi-automated data pipeline. After each dark and flatfield sequence is averaged in time, the gaintable spectrum is calculated as the normalized difference between flatfield and dark. The flatfielded spectral sequence is obtained by subtracting the corresponding dark spectrum from each solar or comb spectrum and dividing by the gaintable spectrum. 

In the next processing steps, the solar spectrum has to be calibrated to an absolute wavelength grid (using the comb spectrum) and corrected from inherent systematic wavelength shifts (calculated by an ephemeris code). Given the strong effect of systematic relative motions between the telescope and the light source, the latter substantially depends on a high accuracy of the applied models. The applied ephemeris code developed by \citet{Doerr2015}, which in turn is based on NASA's Navigation and Ancillary Information Facility Spacecraft Planet Instrument C-matrix Events (SPICE) toolkit \citep{Acton1996}, computes the relative motion between the observing telescope and the Sun with an accuracy of {fractions of} ${\rm mm\,s^{-1}}$. For a ground-based telescope, it includes the orbital motion of the Earth around the Sun, as well as the terrestrial rotation at the location of the observatory, which can add up to a rapidly changing line-of-sight velocity of the order of ${\rm \pm1000\,m\,s^{-1}}$. The next systematic component is the gravitational shift caused by the Sun and Earth according to the General Theory of Relativity which, taken together, amounts to a redshift of $633.3\,{\rm m\,s^{-1}}$, everywhere on the solar disk. If desired, the differential rotation of the Sun reaching a line-of-sight velocity of up to ${\rm \pm2000\,m\,s^{-1}}$ at the solar limb can be modeled and reduced for each heliographic position on the solar disk. The ephemerides code calculates the wavelength shifts for each time step of the observed data sequence. The generated file is later applied to the calibrated solar spectrum.

{To} get the absolute wavelength grid for the solar spectrum, the comb spectrum has to be unambiguously calibrated. The separation of the comb mode is fixed and amounts to 8.0\,GHz. With this knowledge, each pixel of the detector can be assigned to a fraction of a mode number. To determine the positions of the whole emission modes, each individual comb line is fitted with a Gaussian {model with four degrees of freedom (centre, width, amplitude, offset). Due to the very good side-mode suppression by the two high-finesse FPCs, the Gaussians fit the comb line center extremely well. The typical statistical error for the measured mode positions is estimated to $5\,{\rm cm\,s^{-1}}$ for a single calibration exposure. Only potential asymmetries of the instrumental profile can become a severe systematic issue. In this case, the comb lines shift toward the centroid of the instrumental profile. However, the deviation of the final mode profile from the perfectly symmetrical Gaussian fit is minor, so that this effect is limited to a shift well below $1\,{\rm m\,s^{-1}}$ in average. In the next calibration step, the overall} pixel-to-mode-number solution is fitted by a polynomial, thus yielding an intermediate frequency solution. {For a single measurement, its absolute accuracy was determined to about $60\,{\rm cm\,s^{-1}}$. The limiting unmodelled sub-structure in the frequency solution is considered to stem from small (${\rm\sim 50\,nm}$) pixel spacing variations of the CCD detector.}
Since the repetition rate and offset frequency ($-100$\,MHz) of the comb spectrum are known, it is sufficient to unambiguously identify only one mode with its mode number. We select a well-known solar spectral line and enter its air wavelength. The closest comb mode serves as the reference mode to assign all other modes with a well-defined frequency (e.g., in Fig.\,\ref{lars_spectrum} the reference mode close to the \ion{O}{I} line at 6300.3\,\AA\ has a frequency of 475.838\,THz). With this information, all intermediate calibration steps are combined to compute the final frequency (or wavelength) solution. {Each individual solar spectrum is then calibrated by the temporal interpolation of the two adjacent LFC calibration functions.} 
{To achieve a high spectral accuracy for our measurements, a temporal cadence of only few seconds or less is necessary. In fact, the Spectrograph instability introduces the largest error, compared to the line shifts of a few ${\rm cm\,s^{-1}}$ from the LFC and CCD detector. 
Instrument internal seeing, i.e. a spectrograph jitter at the scale of 10\,s, influences the repeatability of the observation. In periods of good instrumental seeing, the error typically is of the order of a few ${\rm cm\,s^{-1}}$ but can increase up to a few ${\rm m\,s^{-1}}$. For measurements with 1\,s cadence, this limits the total accuracy to around ${\rm 1\,m\,s^{-1}}$, or better.}

{Since} the transmission plateau of the spectrograph is quite narrow and the observed spectral band is not always well centered within the diffraction order selected with the predisperser), gradients in the spectral continuum may occur. To correct for this effect, the Fourier Transform Spectrometer \citep[FTS,][]{1999SoPh..184..421N} atlas spectrum is overplotted to the LARS observations. The FTS atlas features a spectral resolution of 400,000 and an accuracy of around ${\rm \pm50\,m\,s^{-1}}$ in the visible which is sufficient for a comparison of both spectra. By selecting individual continuum positions, a polynomial correction function is fitted and stored. In the last step of the data calibration, this correction function and the computed wavelength calibration are applied to the sequence of solar spectra. A final, calibrated LARS spectrum is plotted in Fig.\,\ref{lars_spectrum} as red curve.

%##############################################################################################
\section{First results and outlook}\label{sec_results}
In 2016, three observation campaigns were carried out with the LARS instrument. In total 58 days of telescope time at the VTT were at our disposal to test the instrument with the upgraded LFC and to perform solar observations. All data were calibrated with the processing techniques described in Sect.\,\ref{sec_data}. The main goal was to measure spectral line shifts caused by the solar convection and acoustic waves with an unprecedented spectral accuracy. In a series of forthcoming papers, we will present the results for the center-to-limb variation of the convective blueshift observed with several frequently used spectral lines (e.g., the \ion{Fe}{I} lines at 6301.5\,\AA, 6302.5\,\AA, 6173\,\AA, 5250\,\AA, and \ion{Na}{I} 5896\,\AA). In a further article, we will address umbral oscillations measured with photospheric (\ion{Ti}{I} 5714\,\AA) and chromospheric (\ion{Na}{I} 5896\,\AA) lines. For now, we want to give a brief outlook on the first results and discuss future prospects with LARS.

\begin{figure}[htbp]
\includegraphics[width=\columnwidth]{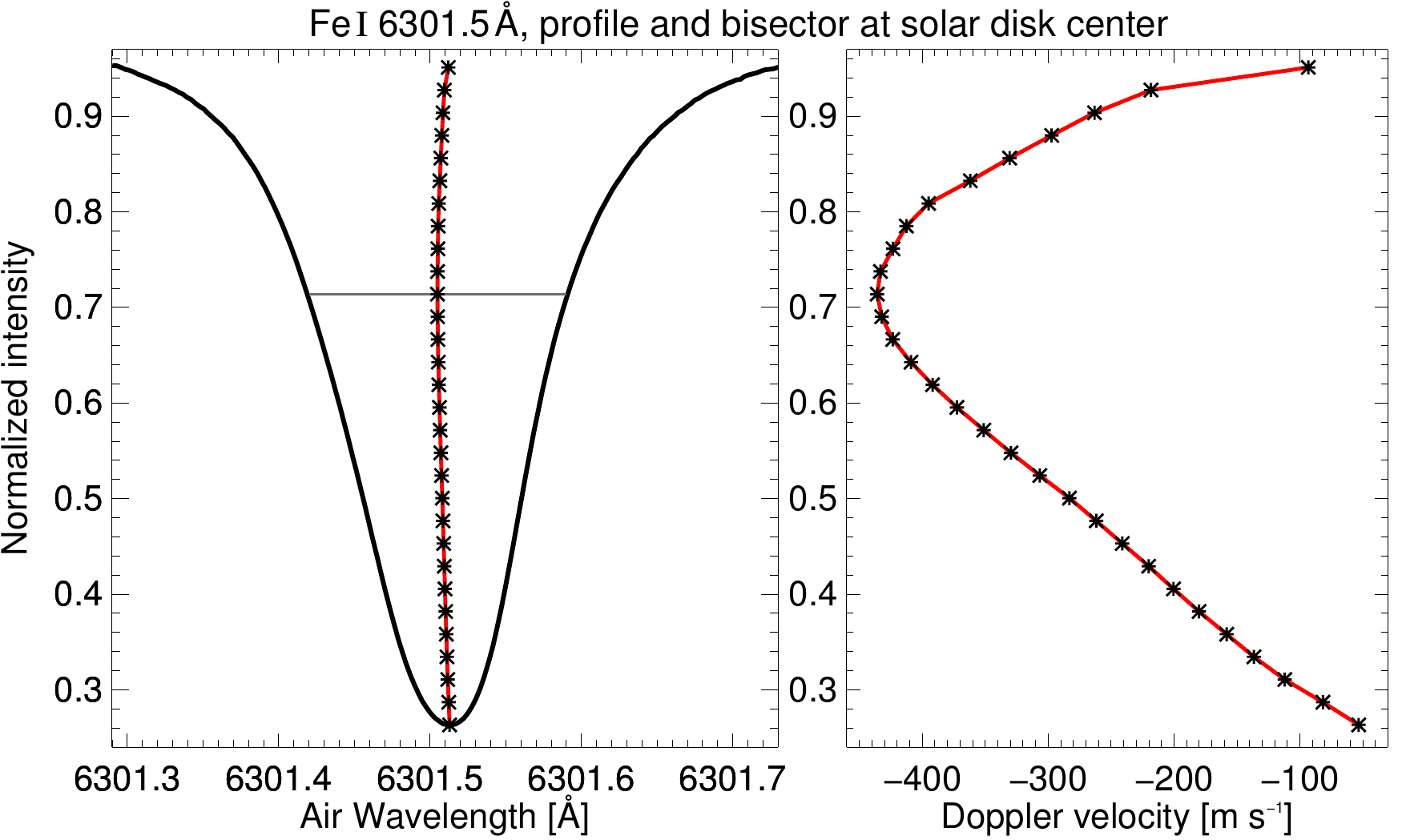}
\caption{Spectral line profile and bisectors of \ion{Fe}{I}\,6301.5\,\AA\ observed {in the Quiet Sun} at disk center. The line profile is drawn as black solid line in the left panel, with intensities normalized to the spectral continuum and air wavelength in \AA. An exemplary bisector is indicated as gray horizontal line. All center positions of the bisectors are shown as black asterisks, their interpolation is displayed as red solid line. In the right panel, the wavelengths were translated into Doppler velocities in  ${\rm m\,s^{-1}}$.}
\label{f_bisector}
\end{figure}

With a spectral resolution ($\Delta\lambda/\lambda$) of 700,000\,--\,800,000 and a spatial sampling of a 1\arcsec--\,10\arcsec\ wide region, LARS enables state-of-the-art spectroscopic investigations of line profiles. In the left panel of Fig.\,\ref{f_bisector}, the intensity profile of the \ion{Fe}{I}\,6301.5\,\AA\ line is shown with its normalized intensity plotted against the air wavelength in \AA. The data was observed on May 13th 2016 at 08:12\,UT with the 10\arcsec\  {fiber-coupling unit in a Quiet Sun region at the solar disk center}. The time series of 20\,min was wavelength calibrated and temporally averaged. The figure is a magnification of the spectral range also shown in Fig.\,\ref{lars_spectrum}. The line has a depth of more than 70\% and a full width at half maximum of around 0.14\,\AA. To attain detailed information on the line shape and Doppler velocities along the sampled atmospheric layers, a bisector analysis was performed on the profile. The center wavelengths are computed for 30 equidistant height levels starting from the line minimum up to 96\% of the continuum intensity. The single positions are marked as black asterisks in Fig.\,\ref{f_bisector}. The interpolated curve was added as solid red line and highlights the asymmetric shape of the line profile. To convert the shifts from air wavelength to Doppler velocities, the exact reference position is required. We typically\footnote{For some spectral lines the reference wavelength positions can been measured with the hollow-cathode lamps available with LARS.} take the observed air wavelength from the atomic spectra database of the National Institute of Standards and Technology \citep[NIST,][]{NIST_ASD}. The reference wavelength is subtracted from the observed wavelength and the remaining difference is translated ($\Delta\lambda/\lambda=\Delta v/c$) into a Doppler shift. The subtraction of the gravitational redshift of $633\,{\rm m\,s^{-1}}$ (uniform for the solar disk) yields the correct line-of-sight velocities in the solar atmosphere. Thanks to the absolute wavelength calibration, there is no need to refer to some (arbitrarily chosen) area of quiet Sun, or the like. At the solar disk center, a redshift is really a downflow, and a blueshift is an upflow in our case.\footnote{This is limited by the accuracy of the reference wavelength, which is better than ${\pm}$1\rm\,m\AA\ (NIST) for the selected well-studied lines.} The right panel in Fig.\,\ref{f_bisector} reflects the bisector analysis from the left panel, plotted against Doppler velocities in ${\rm m\,s^{-1}}$. 

The differential shifts leading to the \glqq C\grqq-shaped bisector curve originate from the physical conditions in the solar atmosphere \citep{1984SoPh...93..219B}. The upflowing material in the centers of granules is hotter than the downward moving gas in intergranular lanes, hence the upflows contribute more to the spatially averaged line profiles, causing a net blueshift of the line. Since the convective flow speed decreases with atmospheric height, the negative (upward) velocities decrease toward the line core. In case of the \ion{Fe}{I}\,6301.5\,\AA\ line, the velocity declines from around $-440\,{\rm m\,s^{-1}}$ at a normalized intensity of $0.7$ to around $-50\,{\rm m\,s^{-1}}$ at the line minimum. 

Beyond this example, the high spectral resolution, wavelength accuracy, and fast temporal cadence of LARS observations enable the detection of subtle and fast changes of the physical properties of the solar atmosphere.

\begin{figure}[htbp]
\includegraphics[width=\columnwidth]{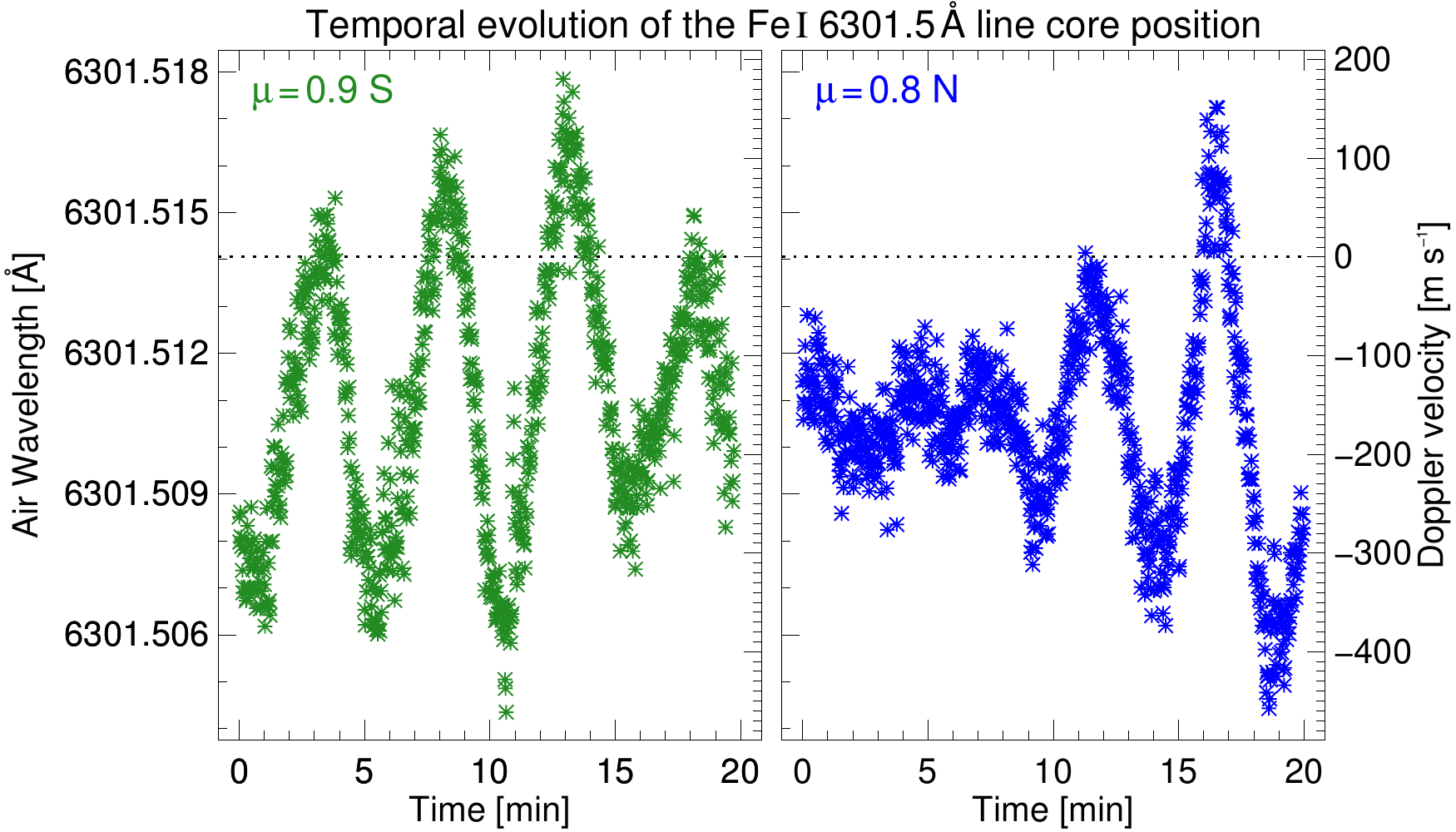}
\caption{Temporal variation of the observed \ion{Fe}{I}\,6301.5\,\AA\ line core position. The oscillatory evolution is displayed for two 20\,min-sequences {observed in the Quiet Sun}. Left panel: Measured on October 14th 2016 at 09:34\,UT at the heliocentric parameter $\mu=0.9$ {on the southern axis of the solar meridian}. Right panel: October 13th 2016 at 15:07\,UT at $\mu=0.8$ on the northern axis {of the meridian}. The single measurements (asterisks) have a cadence of 1.5\,s. The left y-axes scale the wavelength in \AA, the right y-axes are the translated Doppler velocities in ${\rm m\,s^{-1}}$.}
\label{f_pmodes}
\end{figure}

The investigation of acoustic waves in the solar atmosphere with LARS can become invaluable for the research field of local Helioseismology. Magneto-acoustic waves in sunspots and pores can be observed with highest spectral accuracy. Precise measurements of absolute p-mode oscillations can be performed in a confined region of the Sun. Two examples of p-mode observations {in Quiet Sun regions} are presented in Fig.\,\ref{f_pmodes}. The panels show the temporal evolution of the {line center} of \ion{Fe}{I}\,6301.5\,\AA\ in air wavelength (left y-axes) and corresponding Doppler velocities (right axes). {In this case, the line center was defined as the minimum position of a parabolic fit to the inner line core ($\pm20{\rm m\AA}$).} The time sequences were performed with a cadence of 1.5\,s and took 20\,min. Both sequences were measured with the 10\arcsec\ fiber{-coupling unit}. The left {time} series (green asterisks) {was recorded on the southern axis of the solar meridian at an heliocentric angle\footnote{Angle of incidence $\alpha$ between the line of sight and the local normal to the solar surface} of $\alpha=26^{\circ}$, or associated heliocentric parameter $\mu=\cos\alpha=0.9$. The right time series (blue asterisks) was observed at an heliocentric angle $\alpha=37^{\circ}$, or $\mu=0.8$, on the northern meridional axis.} The convective blueshift sets the velocities by $150\,{\rm m\,s^{-1}}$ to negative values. P-mode oscillations with a period of 5\,min and amplitude of up to $300\,{\rm m\,s^{-1}}$ are clearly recognizable in both examples. In addition, a destructive interference of p-mode waves is apparent in the first half of the right series.

3D HD and MHD simulations of the quiet solar atmosphere have been claimed to be \emph{realistic} \citep{Pereira+etal2013} or \emph{highly realistic} \citep{Scott+etal2014b}, and have been used to calibrate the center-to-limb variation of the convective blueshift \citep{delaCruz+etal2011}. LARS provides measurements of unprecedented quality that will allow for a detailed comparison of a number of spectral lines. We will measure the center-to-limb variation of the convective blueshift, and of the corresponding line asymmetries (C-shapes). Since these line properties are intimately coupled with the temperature and velocity gradients in the solar atmosphere, the center-to-limb variations of these quantities will give tight constraints for realistic numerical models of the solar atmosphere. 
As models of the solar atmosphere are fundamental for the interpretation of stellar spectra, such measurements can also have great impact to our understanding of stellar atmospheres. Convective blueshifts on stars other than the Sun are a source of significant systematic errors for the determination of correct radial velocities. Realistic models of stellar atmospheres must faithfully reproduce the observed spectral lines and their center-to-limb variation in the solar atmosphere, before they can be applied to stellar atmospheres. 

The observed temperature of sunspot umbrae suggests the presence of convection, which would also produce some (weak) convective blueshift. To investigate this topic, we intend to determine the absolute wavelength of suitable spectral lines in sunspot umbrae to derive the combined effect of gravitational redshift and convective blueshift. The different center-to-limb behavior of both effects will allow distinguishing between the two contributions. The main limitation for this investigation may come from the rapidly approaching minimum of the current solar activity cycle.

Using the improved possibilities for spectral calibration with laser frequency combs and Fabry-P\'erot interferometers, new solar flux atlases are produced. The atlases of \citet{Molaro+etal2013} using HARPS and of \citet{Reiners+etal2016} using the Fourier transform spectrograph at the Institute for Astrophysics G\"ottingen provide a wavelength accuracy on the order of $\pm10\,{\rm m\,s^{-1}}$, thus exceeding the accuracy of the renowned FTS atlases of \citet{1984sfat.book.....K}, \citet{1999SoPh..184..421N}, or \citet{2011ApJS..195....6W} by one order of magnitude. With LARS, we could construct a disk-center spectral atlas as well. But since the narrow spectral window of the LARS spectrograph does not permit a large wavelength coverage, the fragmented measurement would be very time-consuming and laborious. For mutual benefit, we will provide accurate wavelengths and asymmetries for selected lines of high astrophysical interest to which future studies and atlases can refer to.

%##############################################################################################
\begin{acknowledgements} 
We thank all colleagues at the Kiepenheuer-Institut, at Menlo Systems, and at the Max Planck Institute of Quantum Optics who worked on the development of the instrument, and the upgrade of the laser frequency comb in 2016. We especially thank Thomas Sonner, as well as Roberto Simoes, Frank Heidecke, Andreas Fischer, and Oliver Wiloth for their help in realizing the upgrade of the solar fiber feed optics in 2017. The development of the LARS instrument and the operation of the Vacuum Tower Telescope at the Observatorio del Teide on Tenerife were performed by the Kiepenheuer-Institut f\"ur Sonnenphysik Freiburg, which is a public law foundation of the State of Baden-W\"urttemberg. This work is part of a Post-Doc project funded by the Deutsche Forschungsgemeinschaft (DFG, Ref.-No. Schm-1168/10). The initial astro-comb project for the VTT had been funded by the Leibniz-Gemeinschaft through the "Pakt f\"ur Forschung und Innovation". Finally, we like to thank Reiner Volkmer for his comments on the manuscript.
\end{acknowledgements}
%##############################################################################################

\bibliographystyle{aa} % style aa.bst 
\bibliography{LARS}

\end{document}